\documentclass{article}%
\usepackage{amssymb}
\usepackage{amsmath}
\usepackage{amsfonts}
\usepackage{sw20ssur}
\usepackage{cite}
\usepackage{afterpage}
\usepackage{graphicx}%
\providecommand{\U}[1]{\protect\rule{.1in}{.1in}}

\begin{document}

\title{{\LARGE Social optimality in quantum Bayesian games}}
\author{Azhar Iqbal, James M. Chappell, Derek Abbott\\{\small School of Electrical \& Electronic Engineering, the University of
Adelaide,}\\{\small South Australia 5005, Australia.}}
\maketitle

\begin{abstract}
A significant aspect of the study of quantum strategies is the exploration of
the game-theoretic solution concept of the Nash equilibrium in relation to the
quantization of a game. Pareto optimality is a refinement on the set of Nash
equilibria. A refinement on the set of Pareto optimal outcomes is known as
\emph{social optimality} in which the sum of players' payoffs are maximized.
This paper analyzes social optimality in a Bayesian game that uses the setting
of generalized Einstein-Podolsky-Rosen experiments for its physical
implementation. We show that for the quantum Bayesian game a direct connection
appears between the violation of Bell's inequality and the social optimal
outcome of the game and that it attains a superior socially optimal outcome.

\end{abstract}

\section{Introduction}

The probabilistic approach to the area of quantum games
\cite{MeyerDavid,EWL,EW,Vaidman,BenjaminHayden,EnkPike,Johnson,MarinattoWeber,IqbalToor1,DuLi,Du,Piotrowski,IqbalToor3,FlitneyAbbott1,IqbalToor2,Piotrowski1,Shimamura1,
FlitneyAbbott2,DuNplayer,Hanetal,IqbalWeigert,Mendes,CheonTsutsui,IqbalEPR,NawazToor,OzdemirA,Cheon,Shimamura,ChenWang,Schmidetal,IchikawaTsutsui,CheonIqbal,Ozdemir,FlitneyGreentree,Prevedel,IqbalCheon,IchikawaTsutsuiCheon,Ramzan,FlitneyHollenberg,Aharon,Bleiler,AhmedBleilerKhan,Qiang,FlitneyR,Qiang1,ChappellA,ChappellD,IqbalAbbott,Chappell,IqbalCheonAbbott,Kolenderski,ChappellB,ChappellC,Pawela,Pawela1,Brunner,Frackiewicz,Qiang2,Qiang3,IqbalBayes,Ramanathan,Buscemi,Mohammad,Lason}
shows how quantum probabilities can result in non-classical outcomes that defy
intuition. Without referring directly to the mathematical machinery of quantum
mechanics, this approach shows how a game-theoretic setting can benefit from
quantum probabilities obtained from quantum mechanical experiments. As
classical probabilities, quantum probabilities are normalized but they also
obey other constraints that are dictated by the rules of quantum mechanics.
Our earlier studies \cite{Qiang2,Qiang3} have investigated coevolution
\cite{Perc1} in quantum games and also developing a probabilistic approach to
quantum games \cite{IqbalEPR,IqbalAbbott,IqbalCheonAbbott,IqbalBayes}. The
solution concept of the Nash equilibrium (NE) from non-cooperative game theory
\cite{Binmore,Rasmusen,Osborne} has been studied from the outset of the field
of quantum game theory. Named after John Nash, it consists of a set of
strategies, one for each player, such that no player is left with the
incentive to unilaterally change her action. Players are in equilibrium if the
change in strategies by any one of them would lead that player to earn less
than if she remained with her current strategy.

For a physical system used in implementing a quantum game, one usually
considers violation of Bell's inequality in relation to the outcome of the
game as represented by the concept of a NE. In almost all cases of interest
within the area of quantum game theory, one notices that Bell's inequality
violation may only be indirectly proportional to the NE (or equilibria) of the
game. The question then arises whether there exist other solution concepts for
which the mentioned connection becomes a direct proportionlity. In the present
paper, it is shown that when one considers a particular refinement of the NE
concept, a direct connection can indeed be established between Bell's
inequality and the outcome of the game as represented by that refinement.

Pareto optimality is a refinement on the set of Nash equilibria. An outcome of
a game is Pareto optimal if there is no other outcome that makes every player
at least as well off and at least one player strictly better off. That is, a
Pareto optimal outcome cannot be improved upon without disadvantaging at least
one player. Quite often, a NE is not Pareto optimal implying that players'
payoffs can all be increased. For the Prisoners' Dilemma game, it is well
known that the NE is not Pareto optimal.

A stronger refinement of the NE concept, which is even simpler to state, is
known as \emph{social optimality }\cite{Binmore1}. It is a choice of
strategies, one by each player, which is a social welfare maximizer as it
maximizes the sum of the players' payoffs. Of course, this definition is only
appropriate to the extent that it makes sense to add the payoffs of different
players together. Maximizing the sum of players' payoff may not necessarlily
lead to the satisfaction of all the participating players.

As the Pareto optimal outcome is a refinement on the set of NE, the outcomes
that are socially optimal must also be Pareto optimal. If such an outcome were
not Pareto optimal, there would be a different outcome in which all payoffs
were at least as large, and one was larger. This would be an outcome with a
larger sum of payoffs. On the other hand, a Pareto optimal outcome need not be
socially optimal and a\ NE may not be socially optimal even though it can be
Pareto optimal. However, a socially optimal outcome is always Pareto optimal.

In an earlier work \cite{IqbalBayes} we have studied the solution concept of a
Bayesian Nash equilibrium using a probabilistic approach to quantum games
\cite{IqbalEPR,IqbalAbbott,IqbalCheonAbbott,IqbalBayes}. In the present paper,
we study how quantum probabilities can result in a different outcome for the
solution concept of social optimality in a Bayesian game. We notice that with
this refinement of the NE concept, a direct connection appears between
violation of CHSH form of Bell's inequality \cite{CHSH,ClauserShimony,Peres}
and the social optimal outcome of the Bayesian game.

\section{Social optimality in a Bayesian game}

The decision-makers in a strategic game are called players who are endowed
with a set of actions. In a Bayesian game \cite{Binmore,Rasmusen,Osborne} the
players have incomplete information about the other players' payoffs i.e. the
payoffs are not common knowledge. Incomplete information means that at least
one player does not know someone else's payoffs. Random values are assigned to
the players that take values of \textit{types} for each player and
probabilities are associated to those types. A player's payoff function is
determined by his/her type and the probability associated with that type.
Examples include auctions in which bidders do not know each others' valuations
and bargaining processes in which another player's discout factor is unknown.
The game of Battle of the Sexes \cite{Osborne} when one does not know if the
other prefers to be alone or go on a date provides a clearer exposition of a
social dilemma that can be analysed using the theory of Bayesian games. More
formally \cite{Rasmusen,Osborne}, a Bayesian game consists of players and
states. Each player has a set of actions, a set of signals that the player may
receive, and a signal function that associates a signal with each state.
Consider the following Bayesian game%

\begin{center}
\includegraphics[
height=3.5328in,
width=3.5674in
]%
{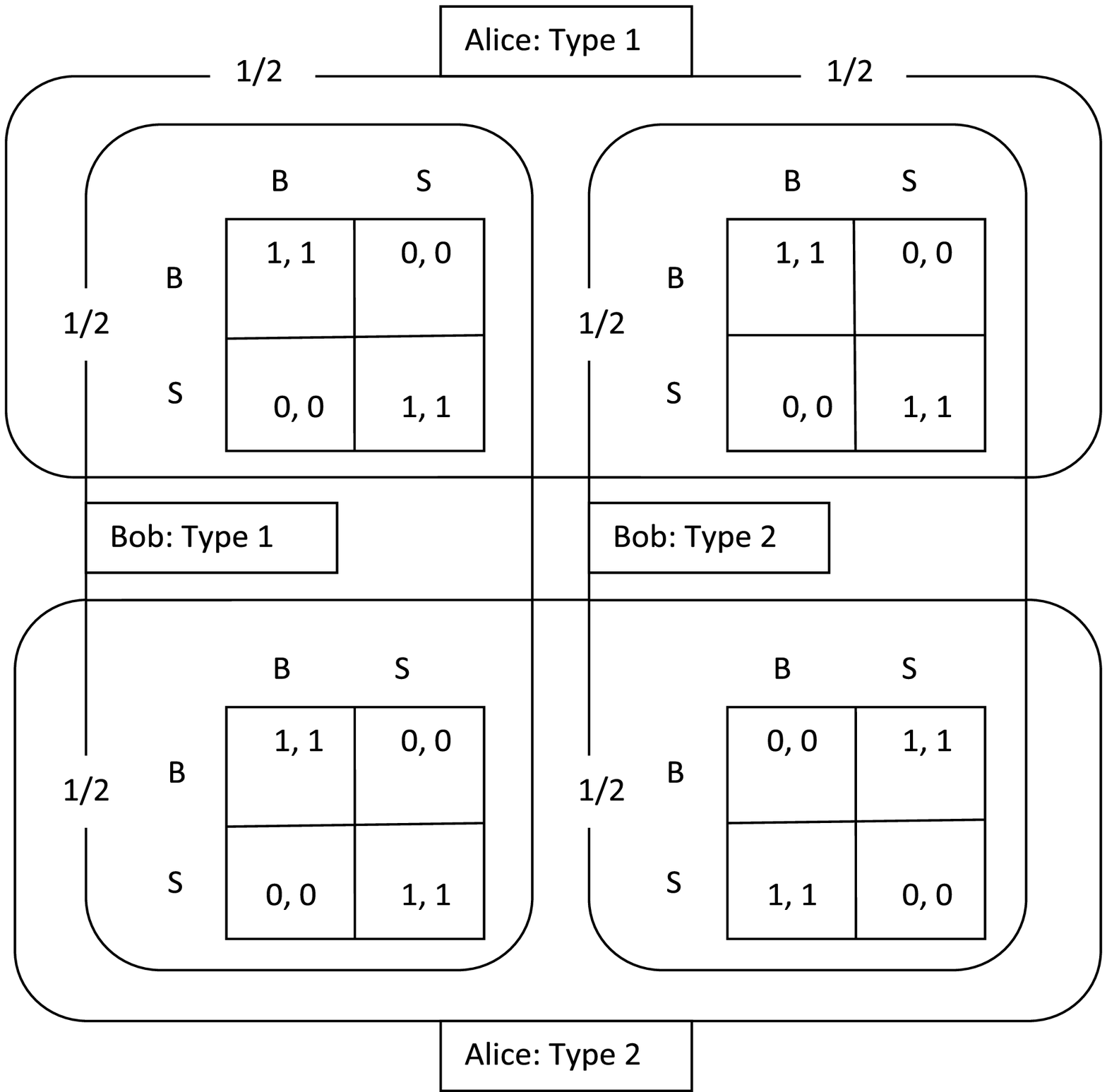}%
\\
Table 1. A Bayesian game of incomplete information.
\label{Fig1}%
\end{center}
in which there are four states represented by four rectangular boxes. As can
be noted, each state is a complete description of one collection of the
players' relevant characteristics. Players do not observe the state and
instead receive signals that provide them information about the state. The
signal functions \cite{Osborne} map states into players' types. For the game
in Table~1, the players Alice and Bob both have two types that we call type
$1$ and type $2$.

As the Table~1 shows, for Alice of either type, Bob is found to be one of the
two types with probability $\tfrac{1}{2}$. Similarly, for Bob of either type,
Alice is found to be one of her two types with the probability $\tfrac{1}{2}$.
The payoffs to Alice and Bob of each type is given in the Table~1. For
instance, from the Table~1. The payoff matrix describing the interaction
between Alice of type $2$ and Bob of type $1$ is given as%

\begin{equation}%
\begin{array}
[c]{c}%
\text{Alice of type }2
\end{array}%
\begin{array}
[c]{c}%
B\\
S
\end{array}
\overset{%
\begin{array}
[c]{c}%
\text{Bob of type }1
\end{array}
}{\overset{%
\begin{array}
[c]{ccc}%
B &  & S
\end{array}
}{\left(
\begin{array}
[c]{cc}%
(1,1) & (0,0)\\
(0,0) & (1,1)
\end{array}
\right)  }}.
\end{equation}
The normal form representation of this Bayesian game is obtained in Table~2 below%

\begin{gather*}%
\begin{array}
[c]{c}%
\text{Strategies for}\\
\text{Alice's two types}%
\end{array}
\overset{%
\begin{array}
[c]{c}%
\text{Strategies for}\\
\text{Bob's two types}%
\end{array}
}{%
\begin{array}
[c]{cccccccc}
& \text{$(B,B)$} &  & \text{$(B,S)$} &  & \text{$(S,B)$} &  & \text{$(S,S)$}\\
\text{$(B,B)$} & (1,\tfrac{1}{2}),(1,\tfrac{1}{2}) &  & (\tfrac{1}%
{2},1),(1,\tfrac{1}{2}) &  & (\tfrac{1}{2},0),(0,\tfrac{1}{2}) &  &
(0,\tfrac{1}{2}),(0,\tfrac{1}{2})\\
\text{$(B,S)$} & (1,\tfrac{1}{2}),(\tfrac{1}{2},1) &  & (\tfrac{1}%
{2},0),(\tfrac{1}{2},0) &  & (\tfrac{1}{2},1),(\tfrac{1}{2},1) &  &
(0,\tfrac{1}{2}),(\tfrac{1}{2},0)\\
\text{$(S,B)$} & (0,\tfrac{1}{2}),(\tfrac{1}{2},0) &  & (\tfrac{1}%
{2},1),(\tfrac{1}{2},1) &  & (\tfrac{1}{2},0),(\tfrac{1}{2},0) &  &
(1,\tfrac{1}{2}),(\tfrac{1}{2},1)\\
\text{$(S,S)$} & (0,\tfrac{1}{2}),(0,\tfrac{1}{2}) &  & (\tfrac{1}%
{2},0),(0,\tfrac{1}{2}) &  & (\tfrac{1}{2},1),(1,\tfrac{1}{2}) &  &
(1,\tfrac{1}{2}),(1,\tfrac{1}{2})
\end{array}
}\\
\text{Table~2. A normal form representation of the Bayesian game of Table~1.}%
\end{gather*}
where, the entries in parentheses on left column are pure strategies for
Alice's two types, respectively. Similarly, the entries in parentheses in the
top row are pure strategies for Bob's two types, respectively. For the two
pairs of payoff entries, the first pair is for Alice's two types and the
second payoff pair is for Bob's two types.

For instance, in Table~2 above consider the entry $(\tfrac{1}{2},1),(\tfrac
{1}{2},1)$ that corresponds when the strategy of Alice's two types is $(S,B)$
and the strategy of Bob's two types is $(B,S)$. Here, understandably, the
first entry in the braces is the strategy of either player of type 1. Now we
refer to Table~1 and notice how Table~2 is obtained from Table~1. At the
intersection of the two rows representing the strategies of $S,$ $B,$ played
by Alice's first and second types, respectively, and the two columns
representing the strategies of $S,$ $B,$ played by Alice's first and second
types, respectively, we find the following entries%

\begin{equation}%
\begin{array}
[c]{c}%
\text{Alice's first type: }S\\
\text{Alice's second type: }B
\end{array}
\overset{%
\begin{array}
[c]{c}%
\begin{array}
[c]{cc}%
\text{Bob's first type: }B & \text{Bob's second type: }S
\end{array}
\end{array}
}{%
\begin{array}
[c]{cccc}%
(0,0) &  &  & (1,1)\\
(1,1) &  &  & (1,1)
\end{array}
},
\end{equation}
from which the payoff to Alice's first type is obtained as $\tfrac{1}%
{2}(0)+\tfrac{1}{2}(1)=\tfrac{1}{2}$ and the payoff to Alice's second type is
obtained as $\tfrac{1}{2}(1)+\tfrac{1}{2}(1)=1$. Similarly, the payoff to
Bob's first type is obtained as $\tfrac{1}{2}(0)+\tfrac{1}{2}(1)=\tfrac{1}{2}
$ whereas the payoff to Bob's second type is obtained as $\tfrac{1}%
{2}(1)+\tfrac{1}{2}(1)=1$. The other payoffs in Table~2 can similarly be
obtained from Table~1.

\subsection{Mixed-strategy version}

Consider now the mixed-strategy version of this game in which the players'
probabilities of selecting $B$ from the pure strategies $B$ \& $S$ are given
by numbers $p,q,p^{\prime},$ and $q^{\prime}$ $\in\left[  0,1\right]  $ for
Alice of type $1$, Alice of type $2$, Bob of type $1$, and for Bob of type $2
$, respectively. The mixed-strategy payoffs for Alice's and Bob's two types is
then obtained from Table~1 as%

\begin{align}
\Pi_{\text{A}_{1}}(p,q;p^{\prime},q^{\prime})  &  =\tfrac{1}{2}\left(
\begin{array}
[c]{c}%
p\\
1-p
\end{array}
\right)  ^{T}\left(
\begin{array}
[c]{cc}%
1 & 0\\
0 & 1
\end{array}
\right)  \left(
\begin{array}
[c]{c}%
p^{\prime}\\
1-p^{\prime}%
\end{array}
\right)  {\small +}\tfrac{1}{2}\left(
\begin{array}
[c]{c}%
p\\
1-p
\end{array}
\right)  ^{T}\left(
\begin{array}
[c]{cc}%
1 & 0\\
0 & 1
\end{array}
\right)  \left(
\begin{array}
[c]{c}%
q^{\prime}\\
1-q^{\prime}%
\end{array}
\right)  {\small ,}\nonumber\\
\Pi_{\text{A}_{2}}(p,q;p^{\prime},q^{\prime})  &  =\tfrac{1}{2}\left(
\begin{array}
[c]{c}%
q\\
1-q
\end{array}
\right)  ^{T}\left(
\begin{array}
[c]{cc}%
1 & 0\\
0 & 1
\end{array}
\right)  \left(
\begin{array}
[c]{c}%
p^{\prime}\\
1-p^{\prime}%
\end{array}
\right)  {\small +}\tfrac{1}{2}\left(
\begin{array}
[c]{c}%
q\\
1-q
\end{array}
\right)  ^{T}\left(
\begin{array}
[c]{cc}%
0 & 1\\
1 & 0
\end{array}
\right)  \left(
\begin{array}
[c]{c}%
q^{\prime}\\
1-q^{\prime}%
\end{array}
\right)  {\small ,}\nonumber\\
\Pi_{\text{B}_{1}}(p,q;p^{\prime},q^{\prime})  &  =\tfrac{1}{2}\left(
\begin{array}
[c]{c}%
p\\
1-p
\end{array}
\right)  ^{T}\left(
\begin{array}
[c]{cc}%
1 & 0\\
0 & 1
\end{array}
\right)  \left(
\begin{array}
[c]{c}%
p^{\prime}\\
1-p^{\prime}%
\end{array}
\right)  {\small +}\tfrac{1}{2}\left(
\begin{array}
[c]{c}%
q\\
1-q
\end{array}
\right)  ^{T}\left(
\begin{array}
[c]{cc}%
1 & 0\\
0 & 1
\end{array}
\right)  \left(
\begin{array}
[c]{c}%
p^{\prime}\\
1-p^{\prime}%
\end{array}
\right)  {\small ,}\nonumber\\
\Pi_{\text{B}_{2}}(p,q;p^{\prime},q^{\prime})  &  =\tfrac{1}{2}\left(
\begin{array}
[c]{c}%
p\\
1-p
\end{array}
\right)  ^{T}\left(
\begin{array}
[c]{cc}%
1 & 0\\
0 & 1
\end{array}
\right)  \left(
\begin{array}
[c]{c}%
q^{\prime}\\
1-q^{\prime}%
\end{array}
\right)  {\small +}\tfrac{1}{2}\left(
\begin{array}
[c]{c}%
q\\
1-q
\end{array}
\right)  ^{T}\left(
\begin{array}
[c]{cc}%
0 & 1\\
1 & 0
\end{array}
\right)  \left(
\begin{array}
[c]{c}%
q^{\prime}\\
1-q^{\prime}%
\end{array}
\right)  {\small ,}\nonumber\\
&  \label{MixedStrategyPayoffs}%
\end{align}
where $T$ is for transpose and the subscripts $1$ and $2$ under A and B refer
to the respective player's type. Also, a semicolon is used to separate Alice's
and Bob's strategic variables $p,q$ and $p^{\prime},q^{\prime}$, respectively.

Referring to how $p,q,p^{\prime},q^{\prime}$ are defined above, the strategies
of the Alice's two types are $(B,B),$ $(B,S),$ $(S,B),$ $(S,S)$ and these can
be written as $(p,q)=(1,1),$ $(1,0),$ $(0,1),$ $(0,0)$, respectively.
Likewise, the strategies of Bob's two types are $(B,B),$ $(B,S),$ $(S,B),$
$(S,S)$ and can be written as $(p^{\prime},q^{\prime})=(1,1),$ $(1,0),$
$(0,1),$ $(0,0)$, respectively. Substituting these in the mixed-strategy
payoffs (\ref{MixedStrategyPayoffs}) the entries of the normal form
representation of the game in Table~2 can then be obtained.

\section{Social optimality in factorizable Bayesian game}

In order to introduce quantum probabilities in this game, we note that
mixed-strategy payoff relations (\ref{MixedStrategyPayoffs}) can be written as%

\begin{align}
\Pi_{\text{A}_{1}}(p,q;p^{\prime},q^{\prime})  &  =\tfrac{1}{2}\left[
(1)\epsilon_{1}+(0)\epsilon_{2}+(0)\epsilon_{3}+(1)\epsilon_{4}\right]
+\tfrac{1}{2}\left[  (1)\epsilon_{5}+(0)\epsilon_{6}+(0)\epsilon
_{7}+(1)\epsilon_{8}\right]  ,\nonumber\\
\Pi_{\text{A}_{2}}(p,q;p^{\prime},q^{\prime})  &  =\tfrac{1}{2}\left[
(1)\epsilon_{9}+(0)\epsilon_{10}+(0)\epsilon_{11}+(1)\epsilon_{12}\right]
+\tfrac{1}{2}\left[  (0)\epsilon_{13}+(1)\epsilon_{14}+(1)\epsilon
_{15}+(0)\epsilon_{16}\right]  ,\nonumber\\
\Pi_{\text{B}_{1}}(p,q;p^{\prime},q^{\prime})  &  =\tfrac{1}{2}\left[
(1)\epsilon_{1}+(0)\epsilon_{2}+(0)\epsilon_{3}+(1)\epsilon_{4}\right]
+\tfrac{1}{2}\left[  (1)\epsilon_{9}+(0)\epsilon_{10}+(0)\epsilon
_{11}+(1)\epsilon_{12}\right]  ,\nonumber\\
\Pi_{\text{B}_{2}}(p,q;p^{\prime},q^{\prime})  &  =\tfrac{1}{2}\left[
(1)\epsilon_{5}+(0)\epsilon_{6}+(0)\epsilon_{7}+(1)\epsilon_{8}\right]
+\tfrac{1}{2}\left[  (0)\epsilon_{13}+(1)\epsilon_{14}+(1)\epsilon
_{15}+(0)\epsilon_{16}\right]  ,\nonumber\\
&  \label{factorizablePayoffs}%
\end{align}
where%

\begin{align}
pp^{\prime}  &  =\epsilon_{1},\text{ }p(1-p^{\prime})=\epsilon_{2},\text{
}(1-p)p^{\prime}=\epsilon_{3},\text{ }(1-p)(1-p^{\prime})=\epsilon
_{4},\nonumber\\
pq^{\prime}  &  =\epsilon_{5},\text{ }p(1-q^{\prime})=\epsilon_{6},\text{
}(1-p)q^{\prime}=\epsilon_{7},\text{ }(1-p)(1-q^{\prime})=\epsilon
_{8},\nonumber\\
qp^{\prime}  &  =\epsilon_{9},\text{ }q(1-p^{\prime})=\epsilon_{10},\text{
}(1-q)p^{\prime}=\epsilon_{11},\text{ }(1-q)(1-p^{\prime})=\epsilon
_{12},\nonumber\\
qq^{\prime}  &  =\epsilon_{13,\text{ }}q(1-q^{\prime})=\epsilon_{14},\text{
}(1-q)q^{\prime}=\epsilon_{15},\text{ }(1-q)(1-q^{\prime})=\epsilon_{16}.
\label{factorizability}%
\end{align}
The set of factorizable probabilities $\epsilon_{i}$ $(1\leq i\leq16)$
satisfies the normalization constraint as described below%

\begin{equation}
\sum_{i=1}^{4}\epsilon_{i}=1,\text{ }\sum_{i=5}^{8}\epsilon_{i}=1,\text{ }%
\sum_{i=9}^{12}\epsilon_{i}=1,\text{ }\sum_{i=13}^{16}\epsilon_{i}=1.
\label{Normalization}%
\end{equation}
To find the Socially Optimal outcome of this game, we consider the sum of
players' payoffs and find its maximum i.e.%

\begin{equation}
\Pi_{\mathrm{sum}}(p,q;p^{\prime},q^{\prime})=\Pi_{\text{A}_{1}}%
(p,q;p^{\prime},q^{\prime})+\Pi_{\text{A}_{2}}(p,q;p^{\prime},q^{\prime}%
)+\Pi_{\text{B}_{1}}(p,q;p^{\prime},q^{\prime})+\Pi_{\text{B}_{2}%
}(p,q;p^{\prime},q^{\prime}). \label{PayoffsSum}%
\end{equation}
From Eqs.~(\ref{factorizablePayoffs},\ref{factorizability}) we can write%

\begin{align}
\Pi_{\text{A}_{1}}(p,q;p^{\prime},q^{\prime})  &  =(p^{\prime}+q^{\prime
})(p-\tfrac{1}{2})+(1-p),\text{ }\Pi_{\text{A}_{2}}(q;p^{\prime},q^{\prime
})=(p^{\prime}-q^{\prime})(q-\tfrac{1}{2})+\tfrac{1}{2},\nonumber\\
\Pi_{\text{B}_{1}}(p,q;p^{\prime},q^{\prime})  &  =(p+q)(p^{\prime}-\tfrac
{1}{2})+(1-p^{\prime}),\text{ }\Pi_{\text{B}_{2}}(p,q;q^{\prime}%
)=(p-q)(q^{\prime}-\tfrac{1}{2})+\tfrac{1}{2},
\end{align}
whose sum can be expressed as%

\begin{equation}
\Pi_{\mathrm{sum}}(p,q;p^{\prime},q^{\prime})=2q^{\prime}(p-q)+2p^{\prime
}(p+q-\tfrac{1}{2})-2p+3.
\end{equation}
Now, an edge solution refers to the end or edge values of the continuous
variables $p,q,p^{\prime},q^{\prime}$ within their allowed intervals that are
$[0,1]$ for each. A `non-edge' solution, however, is the quadruple $(p^{\ast
},q^{\ast};p^{\prime\ast},q^{\prime\ast})$ that maximizes the sum
$\Pi_{\mathrm{sum}}(p,q;p^{\prime},q^{\prime})$. We thus require%

\begin{equation}
\left.  \frac{\partial\Pi_{\mathrm{sum}}}{\partial p}\right\vert _{\ast
}=0,\text{ }\left.  \frac{\partial\Pi_{\mathrm{sum}}}{\partial q}\right\vert
_{\ast}=0,\text{ }\left.  \frac{\partial\Pi_{\mathrm{sum}}}{\partial
p^{\prime}}\right\vert _{\ast}=0,\text{ }\left.  \frac{\partial\Pi
_{\mathrm{sum}}}{\partial q^{\prime}}\right\vert _{\ast}=0,
\end{equation}
where, for instance, $\left.  \frac{\partial\Pi_{\mathrm{sum}}}{\partial
p}\right\vert _{\ast}$ states that this expression is evaluated at $p=p^{\ast
},$ $q=q^{\ast},$ $p^{\prime}=p^{\prime\ast},$ $q^{\prime}=q^{\prime\ast}$. We obtain%

\begin{equation}
p^{\ast}=q^{\ast}=\tfrac{1}{2},\text{ }p^{\prime\ast}=q^{\prime\ast}=\tfrac
{1}{4},
\end{equation}
at which the sum of the payoffs becomes%

\begin{equation}
\Pi_{\mathrm{sum}}(p^{\ast},q^{\ast};p^{\prime\ast},q^{\prime\ast})=\tfrac
{5}{2}. \label{non-edgeSolution}%
\end{equation}
As can be noticed, the `edge values' of the function $\Pi_{\mathrm{sum}%
}(p,q;p^{\prime},q^{\prime})$ can nevertheless achieve higher values than
$\frac{5}{2}$. For instance, for $(p,q;p^{\prime},q^{\prime})=(0,0;0,0)$ we
have $\Pi_{\mathrm{sum}}(p^{\ast},q^{\ast};p^{\prime\ast},q^{\prime\ast})=3$
and for $(p,q;p^{\prime},q^{\prime})=(1,1;1,1)$ we have $\Pi_{\mathrm{sum}%
}(p^{\ast},q^{\ast};p^{\prime\ast},q^{\prime\ast})=4.$

\section{Social optimality in the Bayesian game with quantum probabilities}

The above analysis of mixed-strategy Bayesian Nash equilibria is for the case
when the underlying probabilities are factorizable. We now enquire whether
there exists a different outcome of this Bayesian game when underlying
probabilities are not subjected to the constraints of factorizability as
described by Eqs.~(\ref{factorizability}). For this we consider a set of
quantum probabilities that are obtained from the generalized
Einstein-Podolsky-Rosen (EPR) experiments as such probabilities can become non-factorizable.

The setting of the generalized EPR experiments
\cite{EPR,Bohm,Bell,Bell1,Bell2,Aspect,ClauserShimony,CHSH,Peres,Cereceda}
involves preparation of a maximally entangled quantum state of pairs of
photons. In each run, two halves of an EPR pair originating from the same
source travel in the opposite directions. One half of EPR pair is received by
the observer Alice whereas the other half is received by Bob. Players Alice
and Bob are assumed to be located in space-like separated regions and are
unable to communicate.

In a run, Alice can measure the spin of her half of an EPR pair either along
the $D_{1}$ direction or along the $D_{2}$ direction. In the same run, Bob can
measure the spin of his half of the same EPR pair either along the
$D_{1}^{\prime}$ direction or along the $D_{2}^{\prime}$ direction.

That is, in a run, a directional pair is chosen by the players that is from
$(D_{1},$ $D_{2}^{\prime}),$ $(D_{1},D_{1}^{\prime}),$ $(D_{2},D_{1}^{\prime
}),$ $(D_{2},D_{2}^{\prime})$. An EPR experiment involves rotating a
Stern-Gerlach detector along the chosen directional pair and performing the
quantum measurement. Independent of a directional pair that is chosen by the
players in a run, the outcome of the quantum measurement is either $+1$ or
$-1$ along any measurement direction for both Alice and Bob.

Corresponding to this experiment, one can then obtain a set of $16$
probabilities that can be written as in Table~3 \cite{Cereceda}.%

\begin{align}
&
\begin{array}
[c]{c}%
\text{Alice's two types}%
\end{array}%
\begin{array}
[c]{c}%
\underset{}{%
\begin{array}
[c]{c}%
D_{1}%
\end{array}%
\begin{array}
[c]{c}%
+1\\
-1
\end{array}
}\\
\overset{}{%
\begin{array}
[c]{c}%
D_{2}%
\end{array}%
\begin{array}
[c]{c}%
+1\\
-1
\end{array}
}%
\end{array}
\overset{\overset{%
\begin{array}
[c]{c}%
\text{Bob's two types}%
\end{array}
}{%
\begin{array}
[c]{cc}%
\overset{%
\begin{array}
[c]{c}%
D_{1}^{\prime}%
\end{array}
}{%
\begin{array}
[c]{cc}%
+1 & -1
\end{array}
} & \overset{%
\begin{array}
[c]{c}%
D_{2}^{\prime}%
\end{array}
}{%
\begin{array}
[c]{cc}%
+1 & -1
\end{array}
}%
\end{array}
}}{\left(
\begin{tabular}
[c]{c|c}%
$\underset{}{%
\begin{array}
[c]{cc}%
\varepsilon_{1} & \varepsilon_{2}\\
\varepsilon_{3} & \varepsilon_{4}%
\end{array}
}$ & $\underset{}{%
\begin{array}
[c]{cc}%
\varepsilon_{5} & \varepsilon_{6}\\
\varepsilon_{7} & \varepsilon_{8}%
\end{array}
}$\\\hline
$\overset{}{%
\begin{array}
[c]{cc}%
\varepsilon_{9} & \varepsilon_{10}\\
\varepsilon_{11} & \varepsilon_{12}%
\end{array}
}$ & $\overset{}{%
\begin{array}
[c]{cc}%
\varepsilon_{13} & \varepsilon_{14}\\
\varepsilon_{15} & \varepsilon_{16}%
\end{array}
}$%
\end{tabular}
\ \right)  }.\nonumber\\
& \nonumber\\
&  \text{Table~3. The probabilities in the generalized EPR experiments.}%
\end{align}

\subsection{Probabilities in generalized EPR experiment}

The elements of the probability set $\varepsilon_{j}$ $(1\leq j\leq16)$ are
known to satisfy certain other constraints that are described in the
following. Note that when the directional pair $(D_{1},$ $D_{2}^{\prime})$ is
chosen for all runs of the experiment, the only possible outcomes are
$(+1,+1),$ $(+1,-1),$ $(-1,+1),$ $(-1,-1)$. The same is true for other
directional pairs $(D_{1},D_{1}^{\prime}),$ $(D_{2},D_{1}^{\prime}),$
$(D_{2},D_{2}^{\prime})$. This leads to the normalization constraint%

\begin{equation}%
{\textstyle\sum\nolimits_{j=1}^{4}}
\varepsilon_{j}=1=%
{\textstyle\sum\nolimits_{j=5}^{8}}
\varepsilon_{j},\text{ \ \ }%
{\textstyle\sum\nolimits_{j=9}^{12}}
\varepsilon_{j}=1=%
{\textstyle\sum\nolimits_{j=13}^{16}}
\varepsilon_{j}. \label{normalization}%
\end{equation}
Also, in a particular run of the EPR experiment, the outcome of $+1$ or $-1$
(obtained along the direction $D_{1}$ or direction $D_{2}$) is independent of
whether the direction $D_{1}^{\prime}$ or the direction $D_{2}^{\prime}$ is
chosen in that run. Similarly, the outcome of $+1$ or $-1$ (obtained along
$D_{1}^{\prime}$ or $D_{2}^{\prime}$) is independent of whether the direction
$D_{1}$ or the direction $D_{2}$ is chosen in that run. These requirements,
when translated in terms of the probability set $\varepsilon_{j}$ are
expressed as%

\begin{align}
&
\begin{array}
[c]{cccc}%
\varepsilon_{1}+\varepsilon_{2}=\varepsilon_{5}+\varepsilon_{6}, &
\varepsilon_{1}+\varepsilon_{3}=\varepsilon_{9}+\varepsilon_{11}, &
\varepsilon_{9}+\varepsilon_{10}=\varepsilon_{13}+\varepsilon_{14}, &
\varepsilon_{5}+\varepsilon_{7}=\varepsilon_{13}+\varepsilon_{15},\\
\varepsilon_{3}+\varepsilon_{4}=\varepsilon_{7}+\varepsilon_{8}, &
\varepsilon_{11}+\varepsilon_{12}=\varepsilon_{15}+\varepsilon_{16}, &
\varepsilon_{2}+\varepsilon_{4}=\varepsilon_{10}+\varepsilon_{12}, &
\varepsilon_{6}+\varepsilon_{8}=\varepsilon_{14}+\varepsilon_{16}.
\end{array}
\nonumber\\
&  \label{LocalityConstraint}%
\end{align}

A convenient solution of the system (\ref{normalization},
\ref{LocalityConstraint}) is reported by Cereceda \cite{Cereceda} to be the
one for which the set of probabilities $\upsilon=\left\{  \varepsilon
_{2},\text{ }\varepsilon_{3},\text{ }\varepsilon_{6},\text{ }\varepsilon
_{7},\text{ }\varepsilon_{10},\text{ }\varepsilon_{11},\text{ }\varepsilon
_{13},\text{ }\varepsilon_{16}\right\}  $ is expressed in terms of the
remaining set of probabilities $\mu=\left\{  \varepsilon_{1},\text{
}\varepsilon_{4},\text{ }\varepsilon_{5},\text{ }\varepsilon_{8},\text{
}\varepsilon_{9},\text{ }\varepsilon_{12},\text{ }\varepsilon_{14},\text{
}\varepsilon_{15}\right\}  $ that is given as%

\begin{equation}%
\begin{array}
[c]{l}%
\varepsilon_{2}=(1-\varepsilon_{1}-\varepsilon_{4}+\varepsilon_{5}%
-\varepsilon_{8}-\varepsilon_{9}+\varepsilon_{12}+\varepsilon_{14}%
-\varepsilon_{15})/{\small 2},\\
\varepsilon_{3}=(1-\varepsilon_{1}-\varepsilon_{4}-\varepsilon_{5}%
+\varepsilon_{8}+\varepsilon_{9}-\varepsilon_{12}-\varepsilon_{14}%
+\varepsilon_{15})/{\small 2},\\
\varepsilon_{6}=(1+\varepsilon_{1}-\varepsilon_{4}-\varepsilon_{5}%
-\varepsilon_{8}-\varepsilon_{9}+\varepsilon_{12}+\varepsilon_{14}%
-\varepsilon_{15})/{\small 2},\\
\varepsilon_{7}=(1-\varepsilon_{1}+\varepsilon_{4}-\varepsilon_{5}%
-\varepsilon_{8}+\varepsilon_{9}-\varepsilon_{12}-\varepsilon_{14}%
+\varepsilon_{15})/{\small 2},\\
\varepsilon_{10}=(1-\varepsilon_{1}+\varepsilon_{4}+\varepsilon_{5}%
-\varepsilon_{8}-\varepsilon_{9}-\varepsilon_{12}+\varepsilon_{14}%
-\varepsilon_{15})/{\small 2},\\
\varepsilon_{11}=(1+\varepsilon_{1}-\varepsilon_{4}-\varepsilon_{5}%
+\varepsilon_{8}-\varepsilon_{9}-\varepsilon_{12}-\varepsilon_{14}%
+\varepsilon_{15})/{\small 2},\\
\varepsilon_{13}=(1-\varepsilon_{1}+\varepsilon_{4}+\varepsilon_{5}%
-\varepsilon_{8}+\varepsilon_{9}-\varepsilon_{12}-\varepsilon_{14}%
-\varepsilon_{15})/{\small 2},\\
\varepsilon_{16}=(1+\varepsilon_{1}-\varepsilon_{4}-\varepsilon_{5}%
+\varepsilon_{8}-\varepsilon_{9}+\varepsilon_{12}-\varepsilon_{14}%
-\varepsilon_{15})/{\small 2}.
\end{array}
\label{dependentProbabilities}%
\end{equation}
This allows us to consider the elements of the set $\mu$ as independent variables.

\subsection{Social optimality with quantum probabilities}

Comparing the Table~1 with the Table~3, one notices that the players' types
should correspond to the four directions $D_{1},$ $D_{1}^{\prime},$ $D_{2},$
and $D_{2}^{\prime}$ along which the measurements are performed in the EPR
experiments. We thus associate Alice's first type to the direction $D_{1},$
Alice's second type to the direction $D_{2},$ Bob's first type to the
direction $D_{1}^{\prime},$ and Bob's second type to the direction
$D_{2}^{\prime}$. With these associations, in a run, each of the two
directions $D_{1}$ and $D_{2}$ is chosen with probability $\tfrac{1}{2}$.
Similarly, each of the two directions $D_{1}^{\prime}$ and $D_{2}^{\prime}$
are chosen with the probability $\tfrac{1}{2}$.

This can also be justified as within the considered game the two types of each
player appear with the probability $\tfrac{1}{2}$ which is same as the
probability of selecting between one of the two available directions for each
player in the generalized EPR experiments. As EPR probabilities can become
non-factorizable, it thus motivates investigating how the outcome of the
Bayesian game is affected.

Consider the Table~3 where, for instance, when it is the interaction between
Alice's type $2$ and Bob of type $1$, and the Stern-Gerlach detectors are
rotated along these directions, the probability that both experimental
outcomes are $-1$ is $\varepsilon_{12}$ and the probability that the observer
$1$'s experimental outcome is $+1$ and observer $2$'s experimental outcome is
$-1$ is given by $\varepsilon_{10}$. The other entries in the Table~3 can
similarly be explained.

Relevant to the EPR setting is the CHSH (Clauser-Horne-Shimony-Holt) form of
Bell's inequality that is usually expressed in terms of the correlations
$\left\langle D_{1}D_{1}^{\prime}\right\rangle $, $\left\langle D_{1}%
D_{2}^{\prime}\right\rangle $, $\left\langle D_{2}D_{1}^{\prime}\right\rangle
$, $\left\langle D_{2}D_{2}^{\prime}\right\rangle $. Using the Table~3 the
correlation $\left\langle D_{1}D_{1}^{\prime}\right\rangle $, for instance,
can be obtained as%

\begin{gather}
\left\langle D_{1}D_{1}^{\prime}\right\rangle =\Pr(D_{1}=1,D_{1}^{\prime
}=1)-\Pr(D_{1}=1,D_{1}^{\prime}=-1)\nonumber\\
-\Pr(D_{1}=-1,D_{1}^{\prime}=+1)+\Pr(D_{1}=-1,D_{1}^{\prime}=-1)\nonumber\\
=\epsilon_{1}-\epsilon_{2}-\epsilon_{3}+\epsilon_{4}.
\end{gather}
Expressions for the correlations $\left\langle D_{1}D_{2}^{\prime
}\right\rangle $, $\left\langle D_{2}D_{1}^{\prime}\right\rangle $, and
$\left\langle D_{2}D_{2}^{\prime}\right\rangle $ can similarly be obtained.
The CHSH sum of correlations is given as%

\begin{equation}
\Delta=\left\langle D_{1}D_{1}^{\prime}\right\rangle +\left\langle D_{1}%
D_{2}^{\prime}\right\rangle +\left\langle D_{2}D_{1}^{\prime}\right\rangle
-\left\langle D_{2}D_{2}^{\prime}\right\rangle , \label{CHSH(a)}%
\end{equation}
and the CHSH inequality stating that $\left\vert \Delta\right\vert \leq2$
holds for any theory of local hidden variables.

The set of constraints on probabilities $\epsilon_{i}$ that are imposed by
Tsirelson's bound\emph{\ }\cite{Tsirelson}\emph{\ }state that the quantum
prediction of the CHSH sum of correlations $\Delta$, defined in (\ref{CHSH(a)}%
), is bounded in absolute value by $2\sqrt{2}$ i.e. $\left\vert \Delta
_{QM}\right\vert \leq2\sqrt{2}$. Taking into account \cite{Cereceda} the
normalization condition (\ref{normalization}), the quantity $\Delta$ can
equivalently be expressed as%

\begin{equation}
\Delta=2(\epsilon_{1}+\epsilon_{4}+\epsilon_{5}+\epsilon_{8}+\epsilon
_{9}+\epsilon_{12}+\epsilon_{14}+\epsilon_{15}-{\small 2}). \label{delta}%
\end{equation}
We now refer to the payoffs given in Eqs.~(\ref{factorizablePayoffs}) and find
that these can be expressed as%

\begin{align}
\Pi_{\text{A}_{1}}(\epsilon_{i})  &  =\tfrac{1}{2}[\epsilon_{1}+\epsilon
_{4}+\epsilon_{5}+\epsilon_{8}],\text{ \ \ }\Pi_{\text{A}_{2}}(\epsilon
_{i})=\tfrac{1}{2}\left[  \epsilon_{9}+\epsilon_{12}+\epsilon_{14}%
+\epsilon_{15}\right]  ,\nonumber\\
\Pi_{\text{B}_{1}}(\epsilon_{i})  &  =\tfrac{1}{2}\left[  \epsilon
_{1}+\epsilon_{4}+\epsilon_{9}+\epsilon_{12}\right]  ,\text{ \ \ }%
\Pi_{\text{B}_{2}}(\epsilon_{i})=\tfrac{1}{2}\left[  \epsilon_{5}+\epsilon
_{8}+\epsilon_{14}+\epsilon_{15}\right]  . \label{EPRpayoffs}%
\end{align}
Note that the payoffs (\ref{EPRpayoffs}) are reduced to the payoffs in the
mixed-strategy game (\ref{MixedStrategyPayoffs}) when the probability set
$\varepsilon_{j}$ $(1\leq j\leq16)$ is factorizable in terms of the
probabilities $p,$ $q,$ $p^{\prime},$ $q^{\prime}\in\lbrack0,1]$ as given by
Eqs.~(\ref{factorizability}). The sum of the players' payoffs thus becomes%

\begin{align}
\Pi_{\mathrm{sum}}(\epsilon_{i})  &  =\Pi_{\text{A}_{1}}(\epsilon_{i}%
)+\Pi_{\text{A}_{2}}(\epsilon_{i})+\Pi_{\text{B}_{1}}(\epsilon_{i}%
)+\Pi_{\text{B}_{2}}(\epsilon_{i})\nonumber\\
&  =\left(  \epsilon_{1}+\epsilon_{4}+\epsilon_{5}+\epsilon_{8}+\epsilon
_{9}+\epsilon_{12}+\epsilon_{14}+\epsilon_{15}\right)  .
\end{align}
Using Eq.~(\ref{delta}) we, therefore, obtain%

\begin{equation}
\Pi_{\mathrm{sum}}(\epsilon_{i})=\tfrac{\Delta}{2}+{\small 2},
\end{equation}
and, as the CHSH form of Bell's inequality is $\left\vert \Delta\right\vert
\leq2$, the range of values for $\Pi_{\mathrm{sum}}(\epsilon_{i})$ when CHSH
version of Bell's inequality holds is%

\begin{equation}
{\small 2}\leq\left\vert \Pi_{\mathrm{sum}}(\epsilon_{i})\right\vert
\leq{\small 3}.
\end{equation}
With the underlying EPR quantum probabilities, the socially optimal outcome of
the game is obtained when $\Pi_{\mathrm{sum}}(\epsilon_{i})$ is maximized,
which is now bounded in by a different absolute value i.e.%

\begin{equation}
{\small 2}\leq\left\vert \Pi_{\mathrm{sum}}(\epsilon_{i})\right\vert
\leq({\small 2+}\sqrt{{\small 2}})={\small 3.414}.
\end{equation}
We note that this sum exceeds the sum of $\tfrac{5}{2}$ for the `non-edge'
solution $(p^{\ast},q^{\ast};p^{\prime\ast},q^{\prime\ast})=(\tfrac{1}%
{2},\tfrac{1}{2};\tfrac{1}{4},\tfrac{1}{4})$ of the socially optimal outcome
in the Bayesian game with factorizable probabilities. It also exceeds the
`edge solution' $(p,q;p^{\prime},q^{\prime})=(0,0;0,0)$ at which this sum is
$3$. However, it is less than the other `edge solution' of $(p,q;p^{\prime
},q^{\prime})=(1,1;1,1)$ at which the sum is $4$.

\section{Discussion}

We study the game-theoretic solution concept of social optimality in a
Bayesian game when the underlying probabilities are obtained from generalized
EPR experiments. In a Pareto optimal solution the players cannot change their
choices so as to make one better off while leaving the other at least as good.
A socially efficient outcome is a set of choices that maximizes the sum of the
two players' payoffs. A socially efficient outcome is always a Pareto optimum,
i.e. it is not possible to make one player better off without also making the
other player worse off. Social optimality is thus a refinement on the set of
Pareto-optimal outcomes and consists of the set of players' strategies that
maximizes the sum of their payoffs.

For generalized EPR experiments, CHSH form of Bell's inequality can be
expressed as a constraint on the absolute value of $\Delta$ as defined in
Eq.~(\ref{delta}) in terms of the joint probabilities $\varepsilon_{j}$
$(1\leq j\leq16)$. For the Bayesian game studied in this paper, we find that
when the underlying probabilities are obtained from generalized EPR
experiments, the sum of the players' payoffs becomes directly proportional to
$\Delta.$ This allows us to observe the effect of violation CHSH form of
Bell's inequality on the socially optimal outcome of the game.

We find that with the underlying probabilities being quantum mechanical, a new
socially optimal outcome is obtained for which the maximum of the sum of
players' payoffs exceeds such maximum for factorizable probabilities while
excluding the edge solutions. However, we also note that the edge solutions
for the factorizable case can still exceed the quantum non-edge solution.
Although the quantum set up is different in this paper, one might argue that
the result developed has direct parallels with the work that is reported in
Ref. \cite{FlitneyR}.

From the words `Bayesian' and `quantum' one can imply the quantum Bayesian
program of Caves, Fuchs, and Schack \cite{Caves} aiming to interpret quantum
probabilities as Bayesian probabilities. In this paper, our aim, however, is
different and consists of showing that Bayesian games provide the appropriate
setting where the role of quantum probabilities can be demonstrated directly
and also without recourse to the mathematical formalism of quantum mechanics.

\end{document}